\documentclass[twocolumn]{revtex4-1}

\def\be{\begin{equation}}
\def\ee{\end{equation}}
\def\la{\label}
\def\bea{\begin{eqnarray}}
\def\eea{\end{eqnarray}}

\def\fr{\frac}

\def\ci{\cite}
\def\la{\label}
\def\bib{\bibitem}

\def\lm{\lambda}

\def\le{\left}
\def\ri{\right}

\def\p{\phi}
\def\Op{\Omega_{\phi}}
\def\Om{\Omega_{m}}
\def\Opo{\Omega_{\phi o}}
\def\Omo{\Omega_{m o}}
\def\rm{\rho_m}
\def\rp{\rho_\phi}
\def\rmo{\rho_{m o}}
\def\rpo{\rho_{\phi o}}
\def\G{\Gamma}

\def\Om{\Omega_{m }}
\def\Omo{\Omega_{m o}}

\def\r{\rho}

\usepackage{makeidx}
\usepackage{ctable}
\usepackage{multirow}
\usepackage{graphicx}
\usepackage{grffile}
\usepackage{epstopdf}
\usepackage{subfig}
\usepackage{subfloat}
\usepackage{colortbl}
\usepackage{appendix}
\usepackage{rotating}
\usepackage{array}

\begin{document}

\title { Scalar Field Dark Energy Parametrization}
\author {Axel de la Macorra}
% \email{macorra@fisica.unam.mx}
\affiliation{Instituto de Fisica, Universidad Nacional Autonoma de Mexico, 04510, Mexico, D.F.}

\begin{abstract}

We propose a new Dark Energy parametrization based on the dynamics of a scalar field. We use  an equation of state
$w=(x-1)/(x+1)$, with $x=E_k/V$, the ratio of kinetic energy $E_k=\dot\phi^2/2$ and potential $V$. The eq. of motion gives $x=(L/6)(V/3H^2)$ and with a solution $x=([1+2 L/3(1+y)]^{1/2}-1)(1+y)/2$ where $y\equiv \rm/V$ and  $L\equiv (V'/V)^2 (1+q)^2,\, q\equiv\ddot\p/V'$. Since the universe is accelerating at present time we use the slow roll approximation in which case we have $|q|\ll 1$ and $L\simeq (V'/V)^2$. However, the derivation of $L$ is exact
 and has no approximation. By choosing an appropriate ansatz for $L$ we obtain a wide class of behavior for the evolution of Dark Energy without the need to specify the potential $V$. In fact $w$ can either grow and later decrease, or other way around,  as a function of redshift and it is constraint between $-1\leq w\leq 1$ as for any canonical scalar field with only gravitational  interaction. Furthermore, we  also calculate the perturbations of DE and since the evolution of DE is motivated by the dynamics of a scalar field the homogenous and its perturbations can be used to determine the form of the potential and the nature of Dark Energy. Since our parametrization is on $L$ we can easily connect it with the scalar potential $V(\p)$.

\end{abstract}

%\pacs{98.62.Dm, 95.30.Cq, 98.62.Gq}
%\keywords {Dark Matter, Rotation Curves, Galactic kinematics, Elementary particle, Mass Model, THINGS}

\maketitle

\section{Introduction}

In the last years the study of our universe has received a great deal of attention since
on the one hand fundamental theoretical questions remain unanswered and on the other hand
we have now the opportunity to measure the cosmological parameters with an extraordinary
precision. Existing  observational experiments involve measurement on CMB \ci{wmap7} or large scale structure LSS \ci{LSS}
or supernovae SN1a \ci{SN}, and new proposals are carried out \ci{planck}.
It has been established that our universe is flat and dominated at present time by Dark Energy "DE" and Dark Matter "DM" with
$\Omega_{DE}\simeq 0.73$,  $\Omega_{DM}\simeq 0.27$ and curvature $\Omega_{k}\simeq-0.012$ \ci{wmap7}. However,
the nature and dynamics of Dark Energy is a topic of mayor interest
in the field \ci{DE.rev}. The equation of state "EOS"  of DE is at present
time $w_o \simeq -0.93$ but we do not have a precise measurements of $w(z)$ as a function
of redshift  $z$ \ci{wmap7}. Since the properties of Dark Energy are still under investigation, different DE parametrization have been proposed to help discern on the dynamics of DE \ci{DEparam}-\ci{quint.ax}. Some of these  DE parametrization have
the advantage of having a reduced number of parameters, but they may lack of a physical motivation
and may also be too restrictive. Furthermore, the evolution of DE may not be enough to distinguish between different models
and the perturbations of DE may be fundamental to differentiate them.

Perhaps the best physically motivated candidates for Dark Energy are scalar fields which can be minimally coupled, only via gravity,  to other fluids \ci{tracker, quint.ax} or can interact weakly in interacting Dark Energy "IDE"\ci{IDE,IDE.ax}. Scalar fields have been widely studied in the literature \ci{tracker, quint.ax} and special interest was devoted to tracker fields \ci{tracker} since in this case the behavior of the scalar field $\p$ is very weakly dependent on the initial conditions at a very early epoch and well before matter-radiation equality. In this class of models the fundamental question of why DE is relevant now, also called the coincidence problem, can be ameliorated by the insensitivity of the late time dynamics on the initial conditions of $\p$. However, at present time we are not concern at this stage  with the initial conditions but we work from present redshift  $z=0$ to larger values of $z$ in the region where DE and its perturbations are relevant. In this case the conditions for tracker fields do not necessarily apply. Interesting models for DE and DM have been proposed using gauge groups, similar to QCD in particle physics, and have been studied to understand the nature of Dark Energy \ci{GDE.ax} and also Dark Matter \ci{GDM.ax}.

Here we propose a new DE parametrization based on scalar fields dynamics, but the parametrization of $w$ can be
used without the connection to scalar fields. This parametrization has a reach structure that allows $w$ to have different evolutions and it may grow and later decrease or other way around. We  also determine  the perturbations of DE which together with the evolution of the homogenous part  can single out the nature of DE. With the underlying  connection between the evolution of $w$ and the dynamics of scalar field we could determine the potential $V(\p)$.
The same motivation of parameterizing  the evolution of scalar field was presented  in an interesting paper \ci{huang}. We share the same motivation but we follow a different path which has the same number of parameters but it has a richer structure and it  is easier to extract information on the scalar potential $V(\p)$.

We organized the work as follows:  in Sec.\ref{ow} we give a brief overview of our DE parametrization. In  Sec.\ref{sfd} we present the dynamics of a scalar field and the set up for our DE parametrization  presented in Sec.\ref{sfp}. We calculate the DE perturbations in Sec.\ref{sec.pert} and finally we conclude in Sec.\ref{con}.

\subsection{Overview}\la{ow}

We present here an overview of our  $w$ parametrization. The EOS is
\be\la{wo}
w=\fr{p}{\r}=\fr{x-1}{x+1}.
\ee
with $x\equiv E_k/V$ the ratio of kinetic energy and potential. The
equation of motion of the scalar field gives  (c.f. eq.(\ref{x3})),
\be\la{xo}
x=\fr{\le(\sqrt{1+\fr{2 L}{3(1+y)}}-1\ri)\le(1+y\ri)}{2}.
 \ee
where $ L= (V'/V)^2 A$, $y= \rm/V$  the ratio of matter and $V$ and $A\equiv (1+q)^2, q\equiv\ddot\p/V'$.
Eq.(\ref{xo}) is an exact equation and is valid for any fluid evolution and/or for  arbitrary potentials $V(\p)$.

The aim of our proposed parametrization for $L,y$  is to cover a wide range od DE behavior. Of course other
interesting parameterizations are possible. From the dynamics of scalar fields we know that the evolution of
$w$ close to present time is very model dependent. For example, in the case of $V=V_o\p^{-2/3}$, used as a
model  of DE derived from gauge theory \ci{GDE.ax}, the shape of  $w(z)$ close to present time depends on the initial
conditions and it may grow or decrease as a function of redshift $z$. Of course if we change the potential
the sensitivity on the choice of $V$ and initial conditions will vary a lot. We also know that tracker fields
are attractor solutions  but in most cases they do not give a negative enough $w_o$ \ci{tracker}. The dynamics
of scalar fields with a single potential term, for a wide class of models,
gives  an accelerating universe only if $\lm=V'/V \rightarrow 0$
or to a constant $|\lm|$ with $w=-1+\lm/3$ \ci{quint.ax}. In this class of models the EOS, regardless
of its initial value, goes to a period of kinetic domination where $w\simeq 1$ and later has a steep transition
to $w\simeq -1$, which may be close to present time, and finally it grows to $w_o$ in a very model and initial condition   dependent. Furthermore,
if  instead of having a single potential term we have two competing terms close to present time, the evolution
of $w(z)$ would even be more complicated. Therefore, instead of deriving the potential $V$ from theoretical
models as in \ci{GDE.ax} we propose to use an ansatz for the functions $L,y$ which on the one hand should cover
as wide as possible the different classes of DE behavior with the least number of parameters, without sacrificing
generality, and on the other hand we like to have the ansatz as close as possible to the know scalar field
dynamics. We believe  that using or model will greatly simplify the extraction od DE from the future observational
data. We propose therefore the ansatz (c.f. eq.(\ref{ll}))
 \bea\la{ly}
L=&&L_o+L_1 y^\xi f(a)=L_o+L_1 y_o^\xi   \le(\fr{a^{3\xi w_o}}{1+(a/a_t)^k}\ri)\\
f(a)=&&\fr{1}{1+(a/a_t)^k}=\fr{1}{1+[(1+z_t)/(1+z)]^k}
\eea
where $L_o,L_1$ are free parameters giving $w_o$ and $w_1=w(z \gg z_t)$  at early times,
$f(z)$ is a function that goes from $f(z=0)=1/(1+(1+z_t)^k)$  at $z=0$ to
$f(z\gg 1) =1$ and $z_t $ sets
the transition redshift between $w_o$ and $w_1$ (a subscript $o$ represents present time quantities)
while $k$ the steepness of the transition and  $\xi$ takes only two values $\xi=1$ or $\xi=0$. We
show that a steep  transition of $w$ has a bump in the adiabatic sound speed $c_a^2$ which could be
detected in large scale structure.  Since the universe is accelerating at present time we may take the slow roll approximation where $|q|\ll 1, A\simeq 1$ and $L\simeq (V'/V)^2$. However, the derivation of $L$ in eq.(\ref{xo}) is exact  and has no approximation. We will show in section \ref{sfp} that $w$ can have a  wide
range of behavior and in particular it can decrease and later increasing as a function of redshift
and viceversa, i.e the shape and steepness are not predetermined by the choice of parametrization. Of
course we could use other parametrization since the evolution of $x$ and $w$ in eqs.(\ref{wo}) and (\ref{xo})
are  fully valid. There is also no need to have any  reference  to the underlying scalar field dynamics,
i.e. it is not constraint to scalar field dynamics.
However, it is when we interpret $x\equiv \dot\p^2/2V$ with $ L= (V'/V)^2 A$ and the ratio $y= \rm/V$ that we
connect the evolution of $w$ to the  scalar potential $V(\p)$.

\section{Scalar Field Dynamics}\la{sfd}

We are interested in obtaining a new DE parametrization  inferred from scalar fields.
Since it is derived from the dynamics of a scalar field $\p$ we can also determine its
perturbations which are relevant in large scale structure formation.
We start from the  equation of motion for a canonical scalar field  $\p(t,x)$ with a potential $V(\p)$
in a FRW metric. The homogenous part of $\p$  has an equation of motion
 \be\la{eqm}
 \ddot \p + 3H\dot\p+V'(\phi)=0
 \ee
 where  $V'\equiv dV/d\p$, $H=\dot a/a$ is the Hubble constant, $a$ is the scale factor with $a_o/a=1+z$
 and a dot represents derivative with respect to time $t$. Since we are interested in the
 epoch for small $z$ we only need to consider matter and DE and we have
 \be\la{hh}
3H^2=\rm+\rp
\ee
in natural units $8\pi G=1$.  The energy density $\r$  and pressure $p$
for the scalar field are
\be\la{rp1}
\rp=\fr{1}{2}\dot\p^2+V ,\;\;\; p_\p=\fr{1}{2}\dot\p^2-V
\ee
and the equation of state parameter "EOS" is
\be\la{w}
w\equiv \fr{p_\p}{\rp}=\fr{\dot\p^2/2-V}{\dot\p^2/2+V}=\fr{x-1}{x+1}
\ee
where we have defined   the ratio of kinetic energy and potential energy as
\be\la{x}
 x\equiv \fr{\dot\p^2}{2V}
 \ee
The value of $x$ gives $w$ or inverting eq.(\ref{w}) we have
$ x=(w+1)/(w-1)$. The evolution of  $x(z)\geq 0  $ determines the Dark Energy  $w$ in the range $-1\leq  w\leq 1$.
For growing $x$ the EOS $w$ becomes larger and at $x\gg 1$ one has $w\simeq 1$ while a decreasing $x$ has
$w$ approaching -1 for $x=0$.

In terms of $x$ and $y\equiv \rm/V$we have
 \be\la{rpx}
 \rp=V(x+1),\;\; p_\p=V(x-1),\;\;  \rm=V y
 \ee
and
 \bea\la{opm}
\Op&=&\fr{1+x}{1+x+y},\;\; \;\;\Om=\fr{y}{1+x+y}, \\
3H^2&=&\rm +\rp=V(1+x+y)
\la{h} \eea
We can write
\be\la{x1}
x\equiv \fr{\dot\p^2}{2V}=  \fr{V}{3H^2}\fr{V'^2}{6V^2}\le(1+q\ri)^2=\fr{V}{3H^2}\fr{L }{6}
\ee
using
\be\la{pq}
\dot\p=-\fr{V'+\ddot\p}{3H}=-\fr{V'(1+q)}{3H}
\ee
and we defined
\bea\la{lq}
L\equiv &&\le(\fr{V'}{V}\ri)^2 (1+q)^2=\lm^2 A,\\
 \lm\equiv &&\fr{V'}{V}, \;\; \;\; q\equiv \fr{\ddot\p}{V'},\;\; \;\; A\equiv\le(1+q\ri)^2  \nonumber
\eea
with $q>-1$. Since the r.h.s. of eq.(\ref{x1})  still depends on $x$ through $H$ we use
eq.(\ref{h}) and eq.(\ref{x1})  becomes then
 \be\la{xy}
 x=\fr{L }{6(1+x+y)}
 \ee
which has a simple solution
 \be\la{x3}
 x=\fr{\le(\sqrt{1+\fr{2 L }{3(1+y)^2}}-1\ri)\le(1+y\ri)}{2}.
 \ee
Eq.(\ref{x3}) sets our DE parametrization as a function of $L$ and $y$. For
small $q$ and $A$ is close to one and  the quantity  $L$  gives direct information on
the potential and its derivative.

\subsubsection{Dynamical evolution  of $x$ and $y$}

Differentiating $x$ and $y\equiv\rm/V$   w.r.t. time we get the evolution
\bea\la{dx}
\dot x &= & \fr{\dot\p}{V}\le(\ddot\p-x V'\ri)=\fr{\dot\p V'}{V}\le(q-x\ri) \\
 &= &6Hx\le(\fr{x-q}{1+q}\ri)
\la{dxx}\eea
and
\be\la{dy}
\dot y =  \fr{\dot\rm}{V} - \fr{yV'\dot\p}{V^2}= 3Hy\le(\fr{2x-q-1}{1+q}\ri)
\ee
were we  used eqs.(\ref{x1}), (\ref{pq})  and $\dot\rm=-3H\rm(1+w_m)$ with $w_m=0$.
The dynamical system  of a scalar field with arbitrary potential  $V$ was studied in \ci{quint.ax},
and the critical points with  $\dot x=0$ in eq.(\ref{dxx}) and constant $x$   are
are $\dot \phi=0$ and $\ddot\p-x V'=0$ or equivalently $x=0, x=q$, respectively.
The first case, $\dot\p=0$, implies
$x=0, w=-1$ and a constant $V(\phi)$ with $\Op \rightarrow 1 $. At the same time, eq.(\ref{dy}) gives
$\dot y= -3Hy$ with has a solution $y=y_i(a/ai)^{-3} \rightarrow 0$ and $\Om \rightarrow 0$.  In the second case,
$q=\ddot\phi/V'=x$,  depending on the value  of $q$  the quantities $x=q$ and $w$ will take different constant values, and for $w<w_m=0$ (i.e. $x<1$)  we will have an increasing in $\Op \rightarrow 1$ \ci{quint.ax}. Setting $q=x$
 in eq.(\ref{dy}) we get $\dot y=3Hyw$ with $w=(x-1)/(x+1)$ constant giving a solution
\be\la{yss}
y=y_o\le(\fr{a}{a_o}\ri)^{3w}.
\ee
The critical points $\dot y=0$ in eq.(\ref{dy})  are   $y=0$ and  $q=2x-1$ with $y$ constant.
In the first case we have $\Om=0$ and $\Op=1$ while in the second case eq.(\ref{dxx})
becomes $\dot x=3H(1-x)/x$ with a solution
\be\la{xss}
e^x(1-x)=e^x_o(1-x_o)\le(\fr{a}{a_o}\ri)^{-3}.
\ee
At large values of $a$ the l.h.s. of eq.(\ref{xss}) vanishes, $x\rightarrow 1, q=2x-1\rightarrow 1$, $w \rightarrow 0$  and $\Om,\Op$ are also constant with $\Op=1-\Om$ (for a generic barotropic fluid the critical point would have been $w\rightarrow w_m$ which in our case is $w_m=0$).
Clearly if $x=q$ ar $x=0$ we are at a critical point $\dot x=0$ in eq.(\ref{dxx}) with
constant $x, w$ but for $x\neq q,0 $ the system evolves and $w$ is in general not constant.

However, in the present work we do not want to study the critical points but the evolution of $x$ close to present time when the universe
is accelerating with $x$ close to zero ($w$ close to -1) but not exactly zero  with $\dot\p\neq 0$ and $\ddot\phi\neq xV'$.
We can assume that $\dot V = V'\dot \p < 0$, since we expect $\p$ to roll down the minimum of its potential,
and the second term in eq.(\ref{dx}) is then negative, while $\ddot\p$ can take either sign.
In the region where  $\ddot \p/V'< x $ we have  $\dot x>0$ while for $\ddot\p/V' > x $ we have $\dot x<0$.
If we take what we call a full slow roll defined by  $\ddot \p=0$ and $3H\dot\p=-V'$ then eq.(\ref{dx}) becomes
\be\la{dx2}
\dot x=-(1+z)H x_z= 6Hx^2
\ee
which is positive definite,  i.e. $\dot x \geq 0, x_z\equiv dx/dz \leq 0 $. The solution to eq.(\ref{dx2}) is
\be\la{sx}
x(a)=\fr{x_o}{1+6 x_o Ln(a_o/a)}=\fr{x_o}{1+6 x_o Ln(1+z)}.
\ee
Therefore if the condition $\ddot \p=0$ or $|q=\ddot \p/V' |\ll x$ is satisfied eq.(\ref{sx}) gives
a decreasing function for $x$ as a function of $z$ and therefore $w(z)$ also decreases. However, we
do not expect to be in a full slow roll regime and
when $x$ is small, e.g. $w<-0.9$ one has $x<0.05$, the slow roll condition
$|\ddot \p|\ <  |V'|$ does not imply that   $\ddot \p \ll x V'$ and the sign of $\dot x$ can be positive or negative
depending on the sign and size of $q=\ddot \p /V'$ compared to $x$ and $x$ can either  grow or decrease.
The value of $q$ parameterizes the amount of slow roll of the potential and a full slow roll has
$q=0$ but we expect to be only in an approximate  slow roll regime  with $|q|\ll 1$ and $A\simeq 1$.
We will discuss further the value of $q$ in section \ref{sec.pert}.

\subsection{Evolution}\la{lim}

Taking the differential of eq.(\ref{xy}) we have
\be\la{dxy}
dx=x_y dy +x_L dL= \fr{-x\, dy}{1+2x+y}+\fr{dL}{6(1+2x+y)}
\ee
where the subscript represents a derivative, e.g. $x_y=dx/dy$,
or in terms of derivative w.r.t. the redshift $z$ we have
\be\la{xz}
x_z=x_y y_z + x_L L_z
\ee
with
\bea\la{xl}
x_y &=& - \fr{x}{6(1+2x+y)}\\
x_L &=&\fr{1}{1+2x+y}.
\eea
Clearly $x_L$ is positive definite while $x_y\leq 0$. The derivative of $w$  is
\be\la{wz}
w_z= w_x x_z
\ee
with
\be
w_x=\fr{2x}{(1+x)^2} \geq 0.
\ee
In general we can assume that DE redshifts slower than matter, at least for small $z$,  since DE has $w<0$ and matter $w_m=0$,
so $y=\rm/V$ is a growing function of $z$, i.e. $y_z>0$.  For $L$ constant we see  from eq.(\ref{xz}) that $x_z=x_yy_z<0$ and  $x$  will decrease  and so will $w$.  On the other hand for $y$ constant we have $x_z=x_ydL$ and an increase on $L$ gives a larger
$x$ and $w$. 

We present in the appendix the dynamical equations of $L$ and $q$ and the  limits of $x$ from eq.(\ref{x3}) 
the following: in the limit $L\ll 1$ we have from  eq.(\ref{xy})
with
\be\la{lm1}
x=\fr{L}{6 (1 + y)},\;\;\;  w=-1+\fr{L}{3(1+y)}
\ee
and therefore $x\rightarrow 0$, $w\rightarrow -1$ as $L\rightarrow 0$.
For $y\gg 1$, with $L$ constant,  we have
\be\la{lm2}
x= \fr{L}{6y},\;\;\;  w= -1 +\fr{L}{3y},
\ee
and  again we have $x\rightarrow 0$, $w\rightarrow -1$ as $y\rightarrow \infty$.
For $y\ll 1$, with $L$ constant,  we have
\be\la{lm2b}
x= \fr{L}{6}, \;\;\; w= -1 +\fr{L}{3}
\ee
giving a constant $x$ and $w$. For $L\ll 1$, with $y$ constant,  we have
\be\la{lm2c}
x= \sqrt{\fr{L}{6}}, \;\;\; w= 1 -2\sqrt{\fr{6}{L}}
\ee
giving a constant $x$ and $w$.
Finally, the limit $L/y\rightarrow L_1$ constant with $y\gg 1$
has a constant $x$ and  $w$   with
\be\la{lm3}
x=\fr{L_1}{6}, \;\;\;  w=\fr{L_1-6}{L_1+6}=-1+\fr{2L_1}{6+L_1}.
\ee
As we see from   eqs.(\ref{lm2}), (\ref{lm2b}) and  (\ref{lm3}) all these limits
are obtained from   eq.(\ref{lm1}), i.e. eq.(\ref{lm1}) is then valid for
the limits $L\ll 1$ or $y\gg 1$ or $y\ll 1$. Clearly depending on the choice of $L$ we can 
have a decreasing or increasing $x, w$ as a function of redshift.

As we see from eq.(\ref{lm2}) for $y\gg 1$ (i.e $\rm\gg V$) valid at larger $z$ 
we can estimate the evolution of $L/y$ from eq.(\ref{dy}) and (\ref{dll}) giving
\be\la{dly}
\fr{d(L/y)}{dt}=\fr{3HL}{y}\le(\fr{\dot H}{3H^2}+\fr{1}{1+q}\ri)
\ee
and using 
\be\la{h3h}
\fr{\dot H}{3H^2}=-\fr{1}{2}(1+w\Op)=-\le(\fr{2x+y}{2(1+x+y)}\ri).
\ee
we have $d(L/y)/dt>0$ for $q>(1+x(1+2y+2x))/(1+2y+3x)$.

\subsubsection{Late time attractor Solution}

The evolution of scalar field has been studied in \ci{quint.ax} and the late time attractor for
scalar fields leading to an accelerating universe with $\Op \rightarrow 1$ requires $w < -1/3, L<2 $. In the limit $|\lm| \gg 1$ one has \ci{quint.ax} with $\rm\ll \rp$ and
\be\la{at1}
\fr{\dot\phi^2}{6H^2}=\fr{L}{6}, \;\;\;\;\;\; \fr{V}{3H^2}=1-\fr{L}{6}
\ee
giving
\be\la{at2}
x\equiv \fr{\dot\p^2}{2V} = \fr{L}{6-L}, \;\;\;\;\;\; w= -1+\fr{L}{3}
\ee
If we take the limit $y\ll 1$ and in eq.(\ref{lm1}) we recover $w$ in eq(\ref{at2}). However, for large z we expect
$y$ to increase and eq.(\ref{at2}) would not longer be valid. In this region we should use eq.(\ref{lm1}) or
the full value $x$ from eq.(\ref{x3}) in eq.(\ref{w}).

\section{Scalar Field DE parametrization}\la{sfp}

In order to have an explicit parametrization of Dark Energy we need to either
choose a potential $V(\p)$ or take a parametrization for $L$ and $y$. Of course
if we want to study a specific potential $V$ we would  solve the  equation of
motion in eq.(\ref{eqm}). However, the aim here is to test a wide class of DE models in order
to constrain the dynamics  of Dark Energy form the observational data.
As discussed in sec.\ref{ow} we will propose an anzatz for $L$ and $y$ that covers
the generic behavior of scalar field  leading to an accelerating universe.

For the quantity  $y=\rm/V$ we propose to assume that $V$ redshifts with
an EOS $w_o\equiv w(z=0) <0$, i.e. $V=V_o(a_o/a)^{3(1+w_o)}$, as in eq.(\ref{yss}).
This does \emph{not} mean that
$\rp\propto  (a_o/a)^{3(1+w_o)}$ since the kinetic energy $\dot\p^2/2$,
or equivalently $x$,  may grow faster or slower than $V$ and the EOS for
DE $w(z)$ will be in general different than $w_o$. We   have then
\be\la{yy}
y=\fr{\rm}{V}=  y_o \le(\fr{a}{a_o}\ri)^{3w_o}= y_o (1+z)^{-3w_o}
\ee
where $ \rm=\rmo (a_o/a)^{3}$, $y_o=\rmo/V_o=2\Omo/\Opo(1-w_o)$ using $V_o=\rpo (1-w_o)/2$.
Clearly the function  $y$ is an increasing function of $z$.
From eq.(\ref{lm2})  we know that as long as $L$ is constant (or growing slower than $y$) that
$x$ and $w$ will decrease as a function of $z$.

If we want to have a constant EOS for DE  at early times $z\gg 1$, as for example
matter $w=0$ or radiation $w=1/3$, which
are reasonable behavior for particles, we should choose $L$ proportional to $y$ for large $y$
or $L/y \rightarrow 0$ if we want  $w\rightarrow -1$ at a large redshift. We  then propose to take
\be\la{ll}
L=L_o+L_1 y^\xi f(z)=L_o+L_1 y_o^\xi   \le(\fr{a^{3\xi w_o}}{1+(a/a_t)^k}\ri)
\ee
where we have chosen
\be\la{f}
f(z)=\fr{1}{1+(a/a_t)^k}=\fr{1}{1+[(1+z_t)/(1+z)]^k},
\ee
and  $L_o,L_1,z_t,k$  are free constant parameters. The function $f(z)$ has as a limit $f(z=0)=[1+(1+z_t)^k]^{-1}$,
$f(z=z_t)=1/2$ and   $f(z\gg z_t)=1$.
The parameters $L_o$ and $L_1$ give $w_o$ and $w$ at  large redshift $z \gg z_t$ while the transition epoch from $w_o$ to $w_1\equiv w(z\gg z_t)$
is given by $z_t$ and $k$ sets the steepness of the transition.
The quantity $\xi$ takes the values $\xi=1$ and $\xi=0$ only and we do not consider it as a free parameter
but more as two different ansatze for $L$.

\begin{figure}[h!]
 \includegraphics[width=3in]{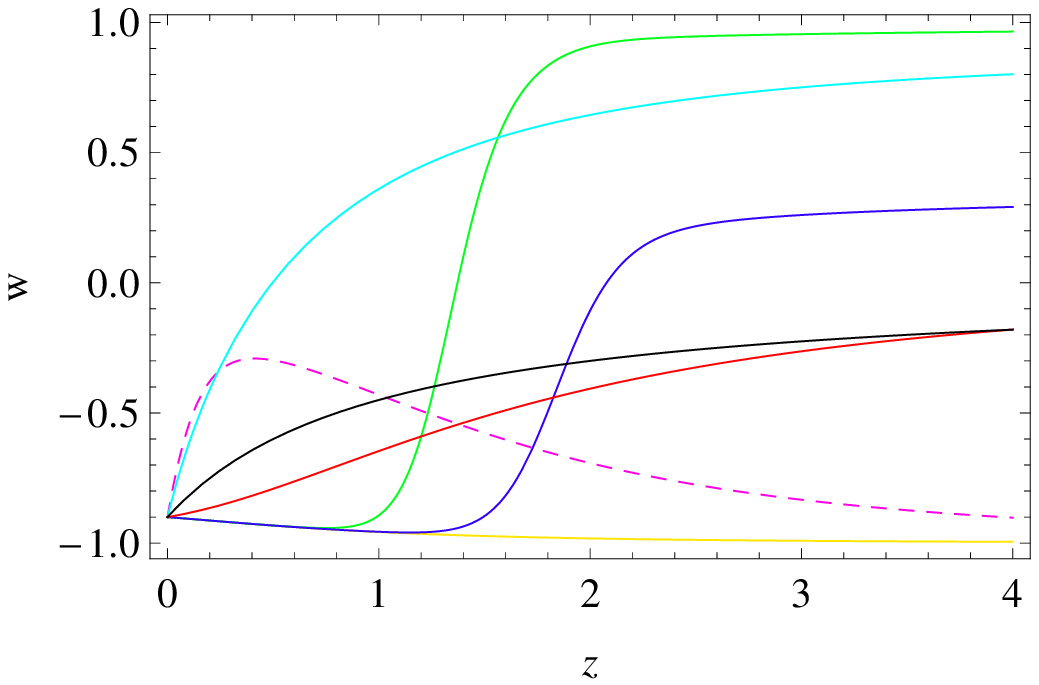}
  \includegraphics[width=3in]{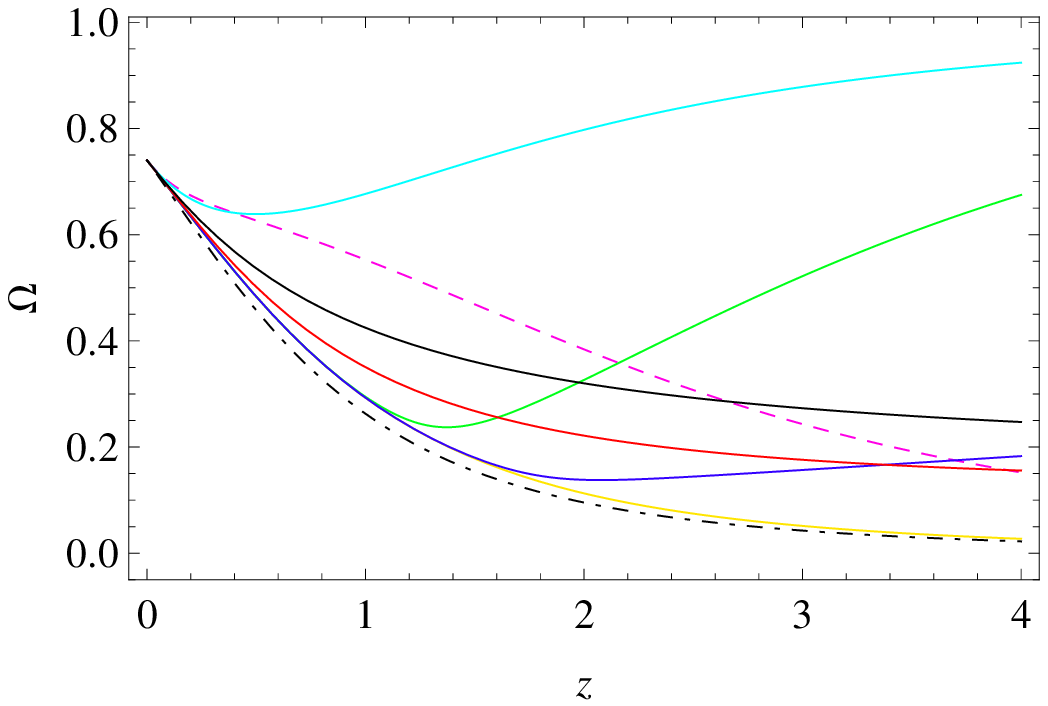}
    \includegraphics[width=3in]{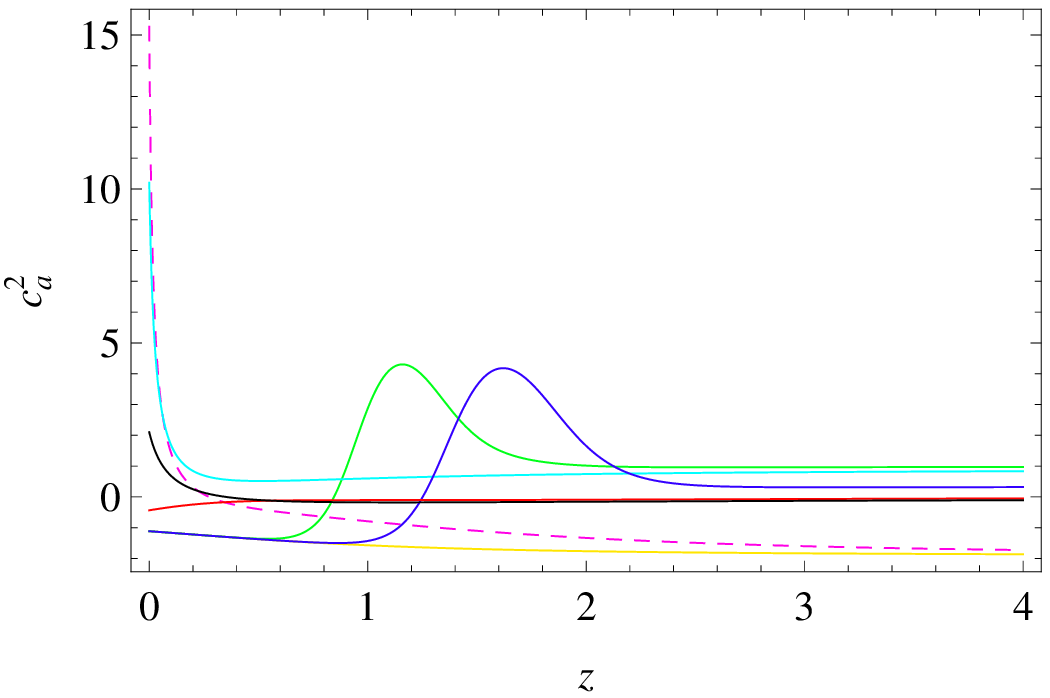}
      \caption{\footnotesize{We show the evolution of  $\Op$, $w$ and $c_a^2$ for different models.  We have taken
    $\xi=1$, $z_t=2$ fixed and $(k,L_1)=(2,6),$$(20,12),(2,100),(20,1000),(20,0)$ (red, dark blue, light blue, green and  yellow, respectively) and $\xi=0$ with  $z_t=0.1, k=10, L1=12$ in pink-dashed line.
   In black we have $w,\Omega_w$ using $w$ in eq.(\ref{w1}) and $\Omega_{cc}$ for a cosmological constant (black dot-dashed).
   We take in all cases $w_o=-0.9$, $\Opo=0.74$ and $L_1=6$ giving $w_1=w(z\gg z_t) =0$ for large $z$. }}
    % z=2, q=2,L=6 rojo,   z=2, q=20,L=12 azul osc,   z=2, q=2,L=100 azul cl,  z=2, q=20,L=100 verde,  z=2, q=20,L=0 amarillo, zt=0.1,q=2,L1=3,xi=0
  \label{FC}
\end{figure}

Using eqs.(\ref{xz}) and  (\ref{xl}) we
can calculate easily $x_z=x_y y_z x_LL_z$   with
\be\la{lz}
L_z =  L_y y_z + L_f f_z=L_1(y_z f + y f_z) ,
\ee
$L_y=\xi L_1 f y^{\xi-1}, L_f=L_1 y^\xi$ and for
\be\la{lyf}
y_z =-\fr{3\xi w_o y}{1+z}, \;\;
f_z =  \fr{k\, f^2}{(1+z)} \le(\fr{1+z_t}{1+z}\ri)^k
\ee
Eq.(\ref{lz}) becomes then
\be\la{lz2}
L_z  =\fr{L_1\, f \,y^\xi}{(1+z)} \le( k\, f \le(\fr{1+z_t}{1+z}\ri)^k   - 3\xi  w_o \ri)
\ee
with  $L_1\, f \,y^\xi=L-L_o$,
or in terms of the scale factor $a$ we have $L_a=(dz/da)L_z= -L_z(a_o/a)^2$  and
\be\la{la2}
L_a  =-\fr{L_1\, f \,y^\xi}{a/a_o} \le( k\, f \le(\fr{a}{a_t}\ri)^k  - 3\xi  w_o \ri)
\ee
For $\xi=1$ eq.(\ref{ll}) allows a wide class of  behaviors for $w$.  If we want $w$ to increase
to $w=0,1/3$ we would take $L_1=6,12$, respectively, or since  in many scalar field models the evolution of $w$
goes form $w_o$ to a region dominated by the kinetic energy density with $w=1$   and in this case we would
should take $L_1 \gg 1$. Of course a
$w(z\gg 1)=1 $ would only be  valid for a limited period  since $\Op$ should not dominated the universe at early times.
We have included in eq.(\ref{ll}) the case $\xi=0$ because we want to allow $w$ to  increase from $w_o$ at small
$z$ and later go to $w\rightarrow -1$ (c.f. pink-dashed line in fig.(\ref{FC})), since this is the behavior of potentials used as a models of DE as for example  $V=V_o\phi^{-n},n=2/3$ derived from gauge group dynamics \ci{GDE.ax} where the behavior
of $w(z)$ close to present time depends on the initial conditions.
% aqui graficas
\begin{figure}[h!]
    \includegraphics[width=3in]{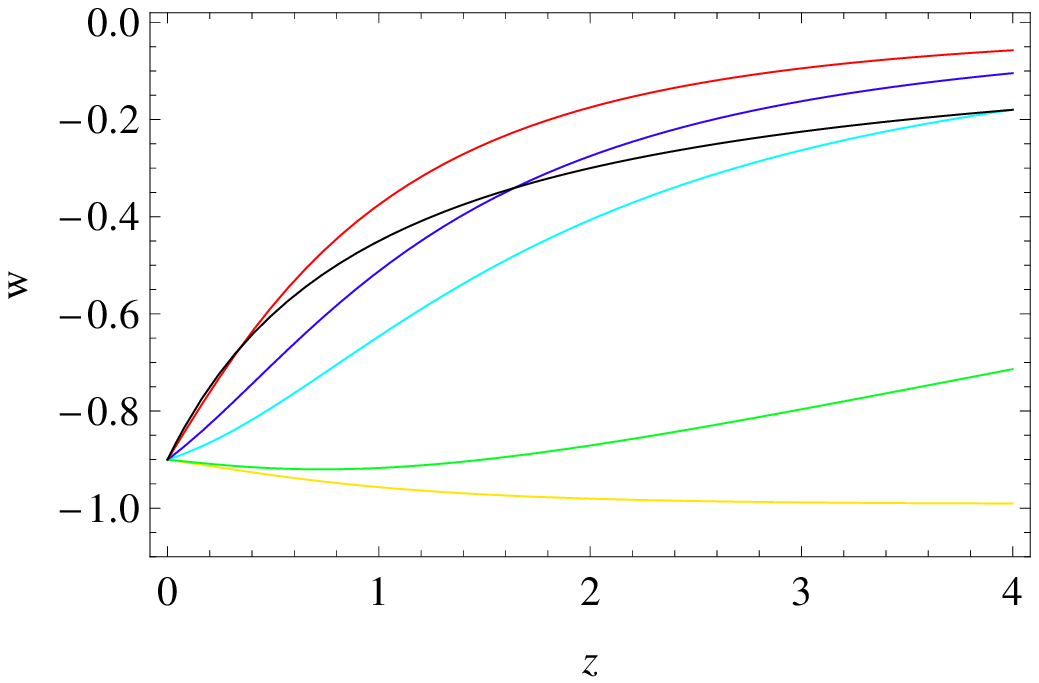}
    \includegraphics[width=3in]{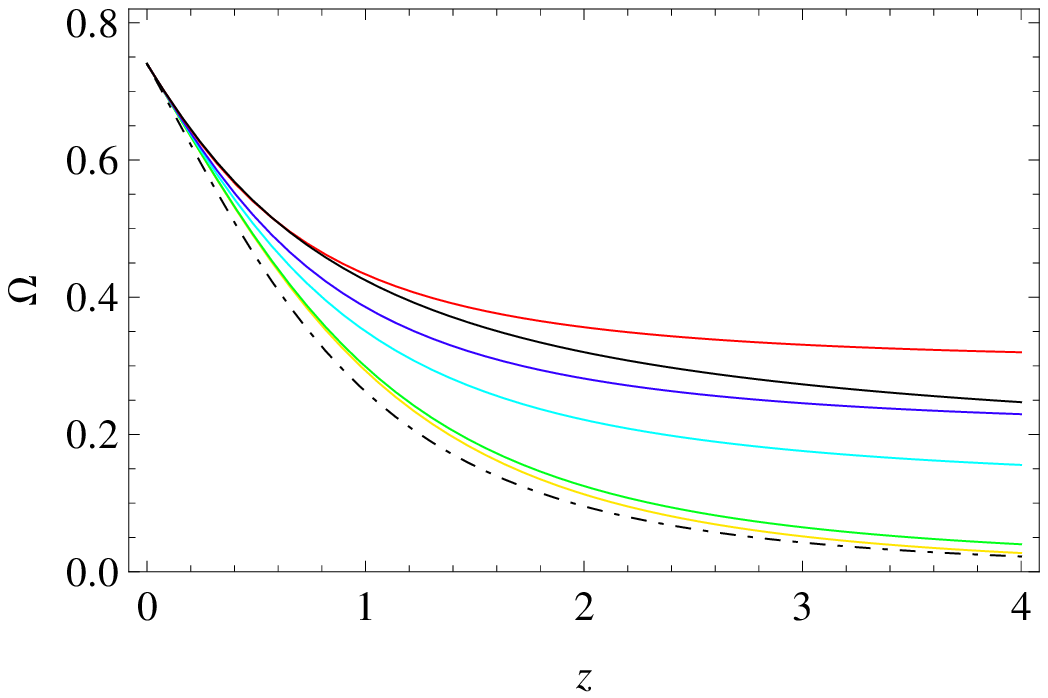}
      \includegraphics[width=3in]{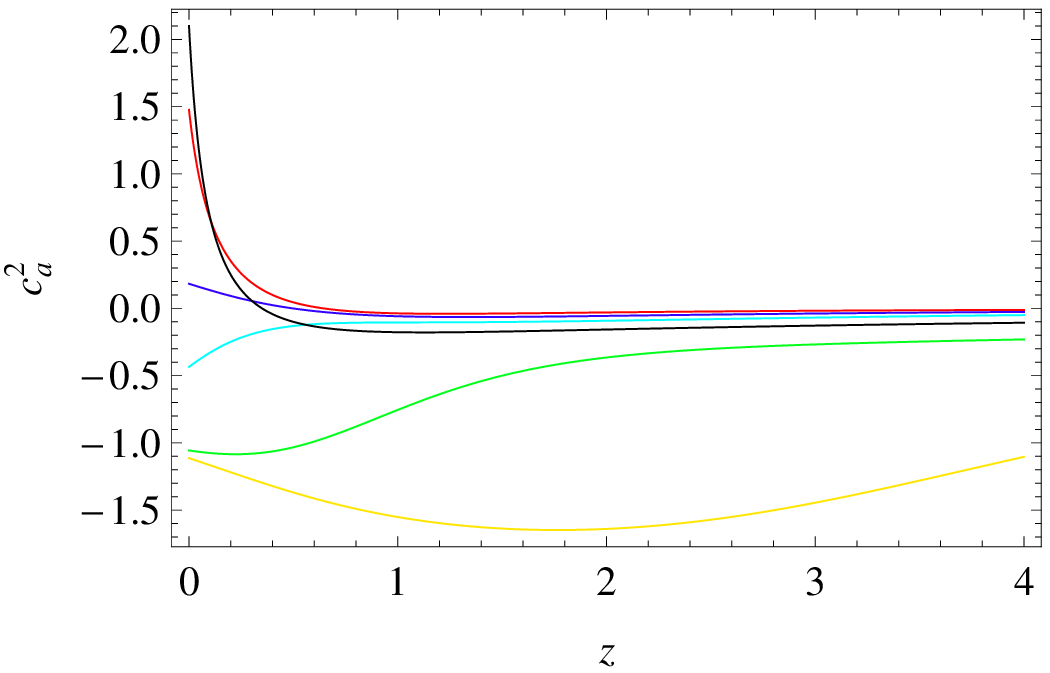}
      \caption{\footnotesize{We show the evolution of  $\Op$, $w$ and $c_a^2$ for different models.  We have taken
    $\xi=1$, $k=2$,$ L_1=6$ with $z_t=0.1,1,$$2,10,100$ (red, dark blue, light blue, green and yellow, respectively).
   In black we have $w,\Omega_w$ using $w$ in eq.(\ref{w1}) and $\Omega_{cc}$ for a cosmological constant (black dot-dashed).  We take in all cases $w_o=-0.9$, $\Opo=0.74$ and $L_1=6$ giving $w_1=w(z\gg z_t)=0$ for large $z$.}}
   % z=.1, q=2,L=6 rojo,   z=1, q=2,L=6 azul osc,   z=2, q=2,L=6 azul cl,  z=10, q=2,L=6 verde,  z=100, q=2,L=6 amarillo
  \label{FA}
\end{figure}
 \begin{figure}[h!]
 \includegraphics[width=3in]{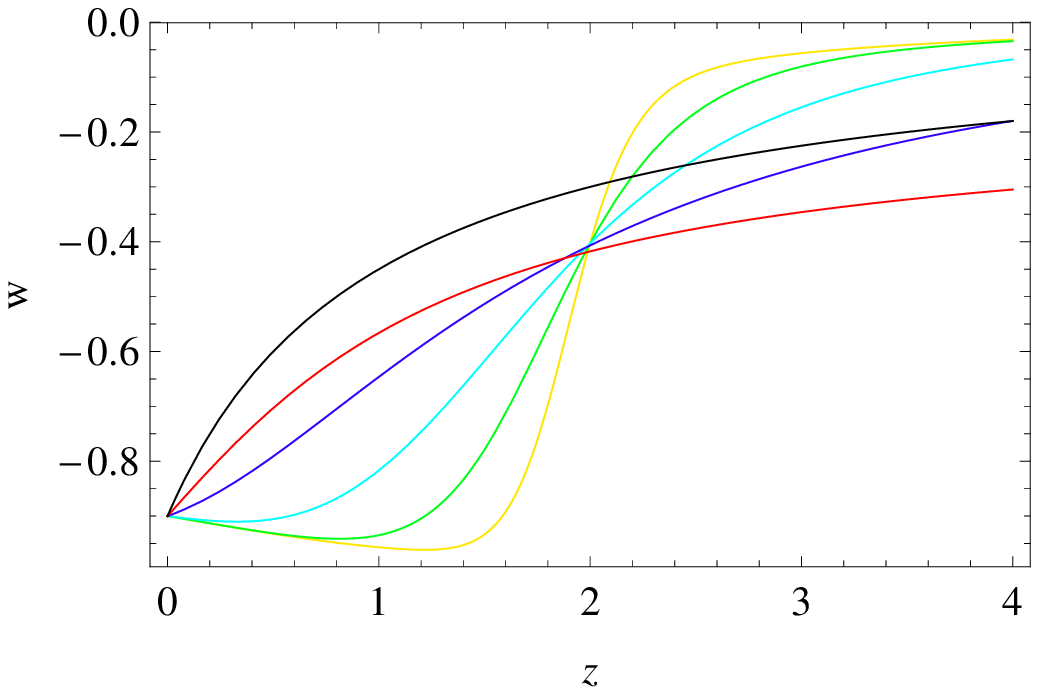}
  \includegraphics[width=3in]{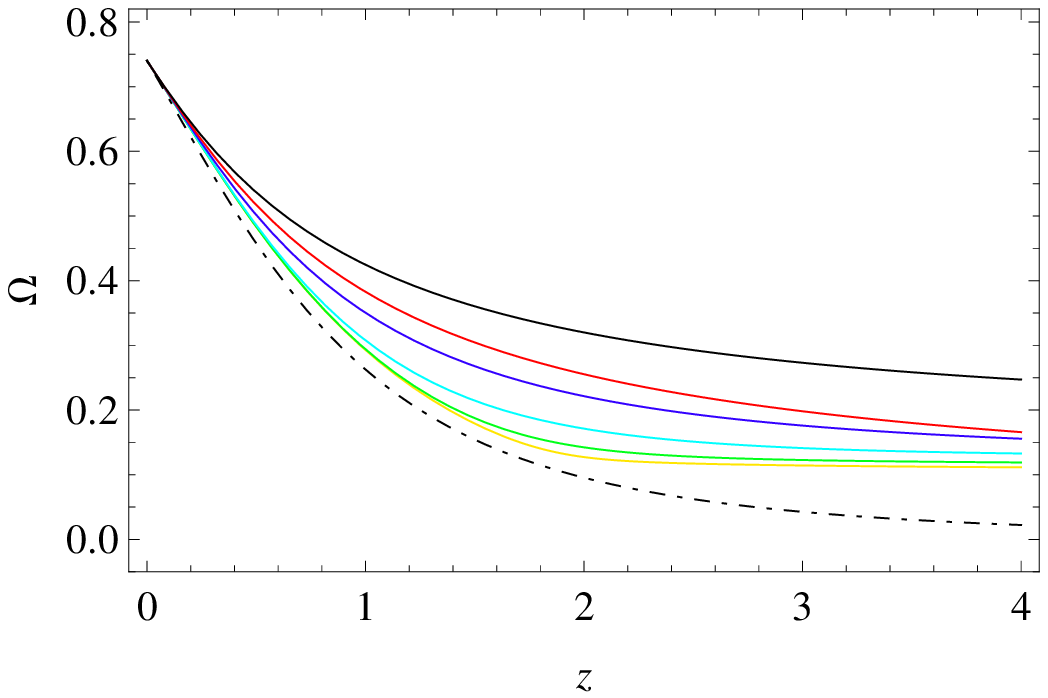}
    \includegraphics[width=3in]{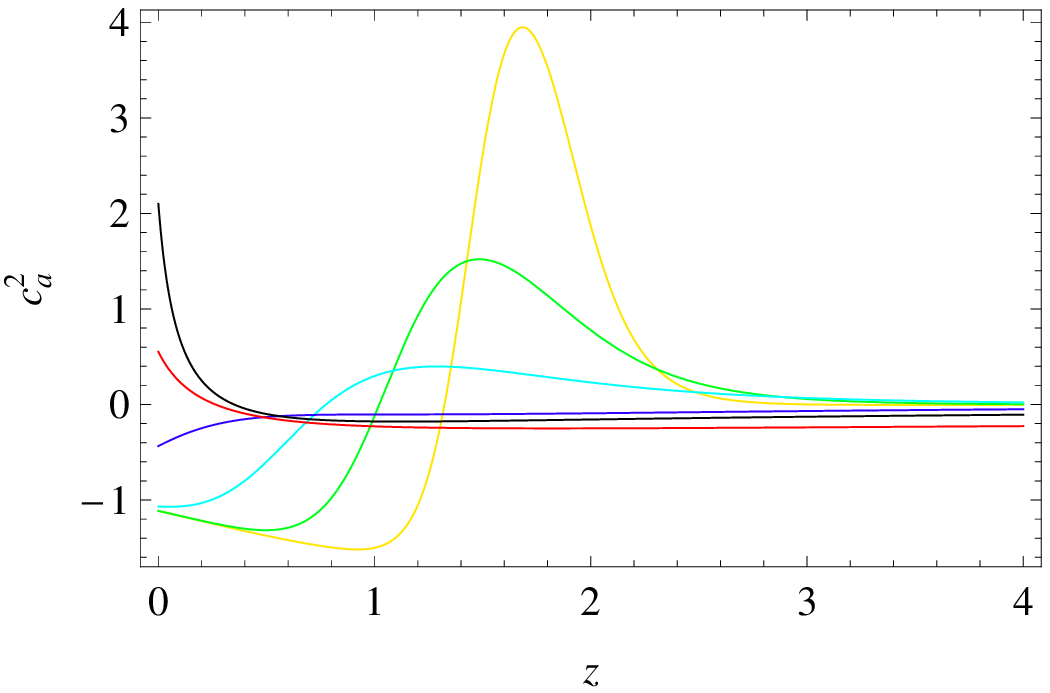}
      \caption{\footnotesize{We show the evolution of  $\Op$, $w$ and $c_a^2$ for different models.  We have taken
    $\xi=1$, $ z_t=2$,$L_1=6$ fixed and $k=1/2,$ $2,$ $5,10,20$ (red, dark blue, light blue, green and  yellow, respectively)
   In black we have $w,\Omega_w$ using $w$ in eq.(\ref{w1}) and $\Omega_{cc}$ for a cosmological constant (black dot-dashed).
   We take in all cases $w_o=-0.9$, $\Opo=0.74$ and $L_1=6$ giving $w_1=w(z\gg z_t) =0$ for large $z$. }}
   % z=2, q=1,L=6 rojo,   z=2, q=2,L=6 azul osc,   z=2, q=5,L=6 azul cl,  z=2, q=10,L=6 verde,  z=2, q=20,L=6 amarillo,
  \label{FB}
\end{figure}

\subsection{Initial Conditions and Free Parameters}

Let us summarize the parameters and initial conditions of our parametrization. The EOS $w$ is only a function of $x$ and
$x$ is a function of $L$ and $y$. From eq.(\ref{yy}) we see that $y$ depends on two parameters $w_o$ and $\Opo$
(or equivalently on $y_o,w_o$), since we are assuming a flat universe with DE and Matter and  $\Omo=1-\Opo$.
From eqs.(\ref{ll}) and (\ref{yy}) we have the initial conditions at present time as
\be\la{yo}
y_o =\fr{\rmo}{V_o}=\fr{2\;\Omo}{(1-w_o)\Opo}
\ee
\be\la{lo}
L_o=\fr{12(1 + w_o) + 6y_o(1 - w_o^2 )}{(1 - w_o)^2}-\fr{L_1y_o^\xi}{1+(1+z_t)^k}.
\ee
For $\xi=1$,  from eq.(\ref{lm3}), the value of $L_1$ gives the early time EOS,  $w_1= w(z\gg z_t) =(L_1-6)/(L_1+6)$ or inverting this expression we have
\be
L_1=\fr{6(w_1+1)}{1-w_1},
\ee
giving for example  $L_1=6$ for $w_1=0$,  $L_1=12$ for $w_1=1/3$,  $L_1=0$ has $w_1\rightarrow -1$
and $L_1\gg 1$  has $w_1\rightarrow 1$. In the case
$\xi=0$ we have the limit $L/y\rightarrow 0$ and  $w\rightarrow -1$, independent on the values
of $L_o, L_1, z_t, q$. From  eq.(\ref{ll}) we have that $L$ depends on  $y$ and $L_o, L_1,z_t,k$ and $w_o,\Opo$.
However, not all parameters are independent, since $L_o$ is a function of $w_o,\Opo,L_1,z_t,q$ and we are left with $\Opo$ and four parameters in $w$.
To conclude, the free parameters are  $\Opo$ and for the EOS we can take   $w_o$, $w_1$, the transition redshift $z_t$
and the steepness of the transition $k$.

\subsection{Other Parameterizations}

We present here some widely used parameterizations and we compare them with our DE present work.
Let is present first a very simple and  widely used DE parametrization  \ci{DELind}
given in terms of only two parameters
\be\la{w1}
w(a)=w_o+w_a(1-a)=w_o+w_a\fr{z}{1+z}
\ee
with a the derivative
\be
\fr{dw}{da}=-w_a, \;\;\;\; \fr{dw}{dz}=\fr{da}{dz}\fr{dw}{da}=(1+z)^{-2}w_a
\ee
Clearly $w$ in eq.(\ref{w1}) is convenient since it is a simple EOS and it has only two parameters.
However, it may be too restrictive  and we do not see a clear connection between the value of
$w$ at small $a$ and its derivative at present time $w_a$. It has only 3 parameters $\Opo,w_o,w_a$,
two less than our model but our model has a much richer structure.

Another interesting parametrization was presented in \ci{corasaniti}. It has 4 free parameters
 \be\la{wg}
 w=w_o+(w_1-w_o)G,\;\;G\equiv \fr{1+e^{a_{d}/d}}{1-e^{1/d}}\fr{1-e^{(1-a)/d_i}}{1+e^{(a_{d}-a)1/d}}
 \ee
 where $w_o,w_2,a_d, d $ are constant parameters. The function $G$ is constraint between $0\leq \G\leq 1$ with
 $G=0$ for $a\gg a_{d}$ and $G=1$ for $a\ll a_{d}$. Therefore
 $a_{d}$ is the scale factor where the transition of the EOS  $w$ goes from $w_o$ to $w_1$
The parameter $d$ gives the width between the transition, for small $d$ the transition from $w_o$ to $w_{1}$ is steeper.  Even though w in eq.(\ref{wg}) gives a large variety of DE behavior \ci{corasaniti}, however the sign of the slope is fixed, and therefore our parametrization in eqs.(\ref{x3}) and (\ref{ll}) has a richer structure
with the same number of parameters.

In the interesting work of \ci{huang} they have followed a similar motivation  as in the present work.
They have presented a DE parametrization motivated by the dynamics of a scalar field.
Their parametrization  has either two three parameters in eqs.(25) and (28) in the paper \ci{huang},
respectively (they do not take $a_{eq}$ as a free parameter but we do think it is an extra parameter).
The two parameter involves the quantities $w_i(a\gg a_o)$, which gives the
EOS at an early time, and $\lm(a_{eq})=V'/V|_{a_{eq}}$  at DM-DE equality (i.e. $\Om=\Omega_{de}$). The second
case,  the parametrization also involves  a term $\zeta_s$ (eq.(23) in \ci{huang}) which depends
on seconde derivative of $V$ and on the value of $\dot\phi/H$ at DM-DE equality.  Since the functional form of
the evolution of the EOS $w(a/a_{eq})$  is fixed in their parametrization the value of $w_o$ at present time is determined if we know  the value of $a_o/a_{eq}$. Therefore, the quantity $a_{eq}$ must also be assumed as
a free parameter.   As in our present work,
they system of equations do not close without the knowledge of the complete $V$ as a function of $\p$. However,
since we are both interested in extracting  information from the observational data to determine the scalar
potential the parametrization given in eqs.(25) and (28) in \ci{huang} is a proposal to study a wide range
of potentials $V$. Here we have taken a different   parametrization which has a closer connection to
the scalar potential $V(\p)$ given by  eqs.(\ref{x}) and (\ref{w}).

\subsection{Results}

We have plotted $w$ for different sets of the parameters in figs.(\ref{FA}),(\ref{FB}) and (\ref{FC})
to show how $w$ depends on $L_o$, $L_1$, $z_t$ and $k$.  We notice that our parametrization in eq.(\ref{ll})
has a very rich structure allowing for $w$ to grow and or decrease at different redshifts. We
have also plotted $\Op$ and the adiabatic sound speed $c_a^2$ defined in eq.(\ref{ca}) for each model.
We are shoing some extreme cases which do not expect to be observational valid but we want to show
the full extend of our parametrization.  We have also include a cosmological constant "C.C." (black dot-dashed) in
the figures of $\Op$ and since $w_{cc}\equiv -1$ and $c_{a\, cc}^2=0$ we do not include them in the  graphs for $w,c_a^2$.
We also plotted  $w,\Op, c_a^2$ (in black) for the parametrization in eq.(\ref{w1}) for comparison.
We take in all cases $w_o=-0.9$, $\Opo=0.74$.

In fig.(\ref{FA}) we show the evolution of  $\Op$, $w$ and $c_a^2$ for different models.
We have taken $\xi=1, k=2,\, L_1=6$ with $z_t=0.1,1,2,10,100$ (red, dark blue, light blue, green and yellow, respectively).    We take in all cases  $L_1=6$ giving $w_1=w(z\gg z_t)=0$. Notice
that  the yellow line the increase to $w=0$  does not show in the graph since $z_t=100$ and we plot $w$ only up  to
$z=4$. The slope in $w$ depends on the value of $z_t$ and since $k=2$ is not large the transition  is not too steep.
The value of $\Op$ decreases slower than a C.C. and for smaller $z_t$ it decreases even more slowly, i.e when $w$ approaches zero faster.

In fig.(\ref{FB}) we show  the evolution of  $\Op$, $w$ and $c_a^2$ for different models. In this
case we have taken $\xi=1, z_t=2,\,L_1=6$ fixed  with $w_1=0$ and we vary $k=1/2,2,5,10,20$ (red, dark blue, light blue, green and  yellow, respectively).
We clearly see in the evolution  $w$ how the steepness of the transition depends $k$ and that $c_a^2$
has a bump at $z_t$ and it is more prominent  for steeper transition. This is generic behavior and we could expect
to see a signature of the transition in large scale structure.

In fig.(\ref{FC}) we show   the evolution of  $\Op$, $w$ and $c_a^2$ and we take
$\xi=1, z_t=2$ fixed and $(k,L_1)=(2,6),(20,12),(2,100),(20,1000),(20,0)$ (red, dark blue, light blue, green and  yellow, respectively)
and $\xi=0$ with  $z_t=0.1, q=10, L1=12$  (pink-dashed line).
In this case  we vary $L_1$  and  we see that for large $L_1$ the EOS $w$
becomes bigger and it may approach $w\simeq 1$ (e.g. green line). Of course this case is not phenomenologically
viable but we plot it to show the distinctive cases of our $w$ parametrization. Once again, a steep transition gives
a bump in  $c_a^2$. The pink-dashed line shows how $w$ can increase at low $z$ and than approach $w=-1$.

We have seen that a our parametrization gives a wide class of $w$ behavior, with increasing and decreasing $w$.
From the observational date we should be  able to fix the parameters of  $L$ in eq.(\ref{ll}) and  we could then
have a much better understanding on  the underlying potential $V(\p)$ using $L=(V'/V)^2$.

\section{Perturbations}\la{sec.pert}

Besides the evolution of the homogenous part of Dark Energy $\p(t)$, its
perturbations $\delta\p(t,x)$ are also an essential ingredient in determining the nature of DE.
The formalism we work  is the synchronous gauge and the  linear perturbations have a  line element
$ds^2=a^2(-d\tau^2+ (\delta_{ij}+ h_{i j})dx^idx^j$, where  $h$ is the trace of the metric perturbations
\ci{ma,hu}. In this sect.(\ref{sec.pert}) a dot represents derivative with respect to conformal time $\tau$
and $\textsf{H}=\dot a/a=(da/d\tau)/a$ is the Hubble constant w.r.t. $\tau$, while $H=(da/dt)/a$

\subsection{Scalar Field Perturbations}

For a DE given in terms of a scalar field, the evolution requires the knowledge of $V$ and
$V'$ while the evolution of  $\delta\p(t,x)$ needs $V''$ through \ci{ma,hu}
\be\la{dp}
\delta \ddot\p + 2\textsf{H} \delta\dot\p+[k^2+a^2V'']\delta\p=-\fr{1}{2}\dot h \dot\p.
\ee
Eq.(\ref{dp}) can be expressed as a function of $a$ with $\dot Y =a\textsf{H} Y_a$ for $Y=\delta\p,\p,\textsf{H}, h$ and  the subscript $a$ means derivative w.r.t. $a$  (i.e. $Y_a\equiv dY/da$), giving
\be\la{dp2}
\delta \p_{aa}+\le(\fr{3}{a}+\fr{\textsf{H}_a}{\textsf{H}}\ri)\p_a +\le[\fr{k^2}{a^2\textsf{H}^2}+\fr{V''}{\textsf{H}^2}\ri]\delta\p=-\fr{1}{2}h_a \p_a.
\ee
In the slow roll approximation we have
\be\la{sl2}
\le|\fr{V''}{3H^2}\ri|=\G x < 3
\ee
where we have used eq.(\ref{x3}) and
\be\la{gt}
\G\equiv\fr{V''V}{V'^2}.
 \ee
 Eq.(\ref{sl2}) implies that an EOS of DE  between $-1\leq w \leq -1/3, 0, 1/3 $, with $0\leq x\leq  1/2,1,2$, requires $\G<3/x=6, 3,3/2$,  respectively.  For a scalar field $\p$ to be in the tracking regime one requires $\G$ to be approximated  constant with $\G>1$ \ci{tracker}. Therefore the regime
$1<\G<3/x$ allows a tracking behavior  satisfying also the slow roll approximation. Here we are more interested in the late time evolution of DE and the tracking regime is not required and in fact we expect deviations from it. However, if $\G$ is nearly constant the evolution of the perturbations in (\ref{dp2})
are then given only in terms of $x$  and we can use our DE parametrization in eq.(\ref{ll}) and eq.(\ref{gt}) and
to calculate them.

We can express the slow roll parameter $\epsilon,  \Upsilon$ in terms of $\G$ and $L$ as
\be
\epsilon \equiv \fr{1}{2}  \le(\fr{V'}{V}\ri)^2 =\fr{L}{2}, \;\;\;\;\;\;  \Upsilon \equiv \fr{V''}{V}=\G L.
\ee
We have decided to use $L,\Upsilon$ instead of $\epsilon,\eta$ not to confuse the reader with the inflation
parameters and the DE ones.

\subsection{Fluid Perturbations}

The evolution of the energy density perturbation $\delta = \delta\r/\r$, $\theta$ the  velocity perturbation
\ci{ma,hu,mukhanov,bean}
\bea\la{p.d.}
 \dot \delta =-(1+w)&&\le(k^2+9\textsf{H}^2\,[c^2_s-c_a^2]\ri)\fr{\theta}{k^2}-\fr{\dot h}{2}\nonumber\\ &&-3\textsf{H}(c^2_s-w)\fr{\delta}{1+w}
 \eea
 \be\la{p.t.}
 \dot\theta=-\textsf{H}(1-3c^2_s)\theta+ c^2_sk^2\fr{\delta}{1+w},
 \ee
and we do not consider an anisotropic stress.
The evolution of the  perturbations depend on three quantities \ci{hu, bean}
\bea\la{cw}
w&=&\fr{p}{\r} \\
\la{ca}c^2_{a}&=&\fr{\dot p}{\dot\r}=w+ \dot w \fr{\rho}{\dot\r}\\
&=&  w - \fr{\dot w }{3\textsf{H}(1+w)}= w+\fr{x_z w_x}{3a(1+w)}\nonumber\\
\la{cs} c^2_{s}&=&\fr{\delta p}{\delta \r}
\eea
where $w$ is the EOS,  $\textsf{H}$ the Hubble constant in conformal time, $c^2_{a}$  is the adiabatic sound speed and
$c^2_{s}$ is the sound speed in the rest frame of the fluid \ci{mukhanov, hu}.
For a perfect fluid one has  $c^2_{s}=c^2_{a}$ but scalar fields are not perfect fluids.
The entropy perturbation  $G_i$ for a fluid $\r_i$ with $\delta_i=\delta \rho_i/\r_i$ are
\be
w_iG_i\equiv (c^2_{si}-c^2_{ai})\delta_i=\fr{\dot p_i}{\dot \r_i}(\fr{\delta p_i}{\dot p_i}-\fr{\delta \r_i}{\dot \r_i})
\ee
where the quantities $G_i$ and $c^2_{ai}$ are scale independent and gauge invariant but $c^2_{si}$
can be neither \ci{ma, bean}. In its rest frame  a scalar filed $\p$ with a canonical kinetic term one
has $c^2_{s}=\delta p/\delta \r=1$ \ci{mukhanov, hu}.  One can relate the rest frame
$\hat{\delta},\hat{\theta}$ to an  arbitrary frame $\delta,\theta$ by \ci{bean}
\be
\hat{\delta}=\delta + 3 \textsf{H}(1+w)\fr{\theta}{k^2}
\ee
and
\be
\delta p=\hat{c}^2_{s}\delta\r+(\hat{c}^2_{s}-c^2_{a})3\textsf{H} (1+w)\r_i\fr{\theta}{k^2}
\ee
As we see from eqs.(\ref{p.d.})-(\ref{p.t.}) the evolution of $\delta$ depends on $c_a^2,c^2_s$ and $w$.
Using eq.(\ref{cw}) and since $w$ is a function of $x(a)$   we have
\be\la{caz}
c^2_a =  w+\fr{x_z w_x}{3a(1+w)}
\ee
with $w_x/(1+w)= 2/[x(1+x)]$. From eqs.(\ref{lyf}) and (\ref{lz2})  we  can
express  $c_a^2$ as a function of the parameters of $x$.

Finally, we can relate $V''$ in terms of the adiabatic sound speed $c^2_a$ in eq.(\ref{ca})  and its time derivative,
using $c^2_a=\dot p/\dot\r=1+2V'/3H(d\p/dt)$, giving
\be\la{cp}
\fr{dc^2_a}{dt}=\le(c^2_a-1\ri)\le(\fr{V''}{V'}-\fr{3H}{2}\le(\fr{2d\dot H}{3H^2}-(c^2_a+1)\ri)\ri)
\ee
and for  $c^2_a\neq 1$ we can invert eq.(\ref{cp}) to give
\be\la{vv}
\fr{V''}{V'}=\fr{\G\,  V'}{V}= \fr{1}{(c^2_a-1)}\fr{dc^2_a}{dt}-\fr{3H}{2}\le(w_T+ c^2_a + 2\ri)
\ee
where we have used $\dot H=dH/dt=-(\r_T+p_T)/2=-3H^2(1+w_T)/2$ with $\r_T,p_T,w_T$ the total energy density,
pressure and EOS, respectively.  In our case we have $\r_T=\rm+\rp,\, p_T=p_m+p_\p=p_\p$ and using
$\rp=V(1+x)$ and eqs.(\ref{w}) and (\ref{h}) we have
\be\la{wT}
w_T\equiv \fr{p_T}{\r_T}= w\Op= \fr{w \rp}{3H^2}=\fr{w(1+x)}{1+x+y}=\fr{x-1}{1+x+y}.
\ee
With eq.(\ref{wT}) the l.h.s. of eq.(\ref{vv}) depends
then only on $y,x$ and are fully determined by our parametrization. In the full slow roll approximation $\ddot\p=0$
and one has $c_a^2=-1$.

\section{Conclusions}\la{con}

We have presented a new parametrization of Dark Energy. This parametrization has a rich structure
and allows for $w(z)$ to have a wide class of behavior, it may grow and later decrease or other way around.
The parametrization of $w$  is given in terms of $x(L,y)$, given in eqs.(\ref{xl}), (\ref{ll}) and
(\ref{yy}). The EOS $w$ is constraint between $-1\leq w\leq 1$  for any value of $x$, with $0\leq x$
by definition. The free parameters of $w$ are $L_o,L_1, z_t, k$, or alternatively
$w_o$ and the EOS at an early time $w_1=w(z\gg z_t)$, given by $L_o$ and $L_1$, respectively (c.f.  eq.(\ref{lo}),
 while $z_t$ gives the transition redshift  between $w_o$ and $w_1$ and
$k$ sets the steepness of the transition. Besides studying the evolution of Dark Energy we also determined
its perturbations from  the adiabatic sound speed $c^2_a$ and $c^2_s$ given in eqs.(\ref{ca}) and (\ref{cs}), which are
functions of $x$ and its derivatives.  We have seen that a steep transition has a bump in  $c^2_a$
and this should be detectable in large scale structure.

We can use the parametrization of $x(L,y)$ in eqs.(\ref{xl}), (\ref{ll}) and (\ref{yy}) and $c^2_a$ and $c^2_s$
in eqs.(\ref{ca}) and (\ref{cs}) without any reference to the underlying physics, namely the dynamics of the scalar
field $\p$, and the parametrization is well defined.  However, it is when we interpret $x=\dot\p^2/2V$ and $ L= (V'/V)^2 A$ and $y= \rm/V$ that we are  analyzing the evolution of a scalar field $\p$ and we can  connect the evolution of $w$ to the  potential $V(\p)$, once
the free parameters are phenomenological determined by the cosmological  data (only when we  take $|q|\gg1, A\simeq 1$
are we taking in the slow roll approximation).

To conclude,  we have  proposed  a new parametrization of DE  which has a rich structure, and
the determination of its  parameters  will help  us to understand the nature of Dark Energy.

\acknowledgments

We acknowledge financial support from  Conacyt Proyect 80519.

\begin{appendix}

\section{}

The parameter $|q\equiv\ddot \p/V'|$ is clearly smaller than one in the slow roll regime ($\ddot\p < 3H\dot\p\simeq V'$), and let us now determine
the dependence of $q$ on the potential $V(\p)$ and its derivatives.
The evolution of $q$ is
\bea\la{dqq}
\dot q &=&\fr{\dddot\p}{V'}-\fr{\ddot \p\dot\p}{V'^2} \\
&=& 3H\le(-q+(1+q)\fr{\dot H}{3H^2}+ 2\G x)\ri)
\eea
where we used eq.(\ref{pq}),
\bea\la{ddd}
\fr{\dddot\p}{V'} &&= - \fr{V''\dot\p}{V'}- \fr{3 \dot H  \dot\p}{V'}-\fr{3H \ddot \p}{V'}\\
&&=-3Hq+3H(1+q)\le(\fr{V''}{9H^2}+\fr{\dot H}{3H^2}\ri),
\eea
and
\be\la{sl3}
\fr{V''}{9H^2}=\fr{\G x }{3},  \;\;\;\;\G\equiv\fr{V''V}{V'^2}.
\ee
We can estimate the value of  $q$ 
 if  we drop  the term proportional to $\dddot\p$ in  eq.(\ref{ddd}) and using eq.(\ref{h3h}) giving
\be\la{qq}
q \simeq \fr{\fr{V''}{9H^2}+ \fr{\dot H}{3H^2}}{1- (\fr{V''}{9H^2}+ \fr{\dot H}{3H^2})}.
\ee
In a stable evolution of $\p$ we have a positive $V''$ and since $\dot H$ is negative
both terms have opposite signs, but of course we do not expect a complete cancelation of these terms.
However both of them are smaller than one, since
$ 0 \leq - \dot H/3H^2<1/2$  for  $x<1$  and
 $|V''/9H^2|=\G x /3 < 1/3$ in the slow roll approximation.
 A tracker behavior requires $\G>1$  \ci{tracker} and
$x<1/\G<1$. Finally, the evolution of $L$ is given by
$\dot L =  2 \lm\dot\lm (1+q)^2+\lm^2 \dot q(1+q)=2L[\dot\lm/\lm+\dot q(1+q)]$,
\bea\la{dll}
\dot L  &=&  \fr{12HLx(1-\G)}{(1+q)}+\fr{2L\dot q}{(1+q)}= 3HL \le(\fr{2x-q}{1+q}+\fr{\dot H}{3H^2}  \ri)\nonumber\\
 &=&  3HL\le(\fr{2(q-x)(1+2x+2y)-y(1+q)}{2(1+q)(1+x+y)}\ri).
\eea
We see that at $-1<q\leq x$ we have $\dot L<0$ giving a decreasing $L$ as a function of time or
an increasing $L$ as  a function of $z$. For $(q-x)/(1+q)>y/(1+2x+2y) $ or
equivalently for $q > (y+2x(1+2x+2y))/(2+4x+3y)$ we have $\dot L >0$  and
a decreasing $L$ as  a function of $z$. The evolution of $d(L/y)/dt$ is given in eq.(\ref{dly}).

Instead of choosing a DE parametrization as in eq.(\ref{ll}) we could solve eqs.(\ref{dqq}) and (\ref{dll})
for different  potentials $V(\p)$ or by taking different approximated solutions or ansatze for
$\G$. However, we choose to parameterize directly $L$ as in eq.(\ref{ll}). Still
using $\dot L= \dot a L_a= a H L_a$ and from eqs.(\ref{la2}),  (\ref{dll}) and $L_1y^\xi f=L-L_o$ we identify
\be
(L-L_o) \le( k\, f \le(\fr{a}{a_t}\ri)^k  - 3\xi  w_o \ri)=  3L \le(\fr{2x-q}{1+q}+\fr{\dot H}{3H^2} \ri)
\ee
and the choices of $\G$ and $q$ would fix the parameters $L_o, k$ and $a_t$.

\end{appendix}

\thebibliography{}

\footnotesize{

 \bib{wmap7}
 E. Komatsu et al., arXiv:1001.4538 [Astrophys. J. Suppl.
Ser. (to be published)];  C. L. Reichardt et al., Astrophys. J. 694, 1200 (2009);
 S. Gupta et al. (QUaD Collaboration), Astrophys. J. 716,
1040 (2010).

\bib{LSS}
B. A. Reid et al., Mon. Not. R. Astron. Soc. 404, 60
(2010);  W. J. Percival et al., Mon. Not. R. Astron. Soc. 327, 1297
(2001); M.Tegmark \textit{et al.}
%Cosmological Constraints from the SDSS Luminous
%Red Galaxies. By SDSS Collaboration (Max Tegmark et al.).
Phys.Rev.D 74:123507 (2006), arXiv:astro-ph/0608632; K. Abazajian \textit{et al.}
 Astrophys.J.Supp.182:543-558,2009.  arXiv:0812.0649v2 [astro-ph]

\bib{SN}
Perlmutter et al., Astrophys. J. 517, 565 (1999); A. G.
Riess et al., Astron. J. 116, 1009 (1998); R. Amanullah et al., Astrophys. J. 716, 712 (2010).

\bib{planck}
%Planck [26] and far future CMBPol [27] experiments.
Planck Collaboration, arXiv:astro-ph/0604069;
 J. Bock et al. (EPIC Collaboration), arXiv:0906.1188.

\bib{DE.rev}
E. J. Copeland, M. Sami,
and S. Tsujikawa, Int. J. Mod. Phys. D 15, 1753 (2006).

\bib{DEparam}

M. Doran and G. Robbers, J. Cosmol. Astropart. Phys. 06 (2006) 026;
E.V. Linder, Astropart. Phys. 26, 16 (2006);
D. Rubin et al. , Astrophys. J. 695, 391 (2009) [arXiv:0807.1108];
J. Sollerman et al. , Astrophys. J. 703, 1374 (2009)[arXiv:0908.4276];
M.J. Mortonson, W. Hu, D. Huterer, Phys. Rev. D 81, 063007 (2010)  [arXiv:0912.3816];
S. Hannestad, E. Mortsell JCAP 0409 (2004) 001 [astro-ph/0407259];
H.K.Jassal, J.S.Bagla, T.Padmanabhan, Mon.Not.Roy.Astron.Soc. 356, L11-L16 (2005);
S. Lee,  Phys.Rev.D71, 123528 (2005)

\bib{DEParam.Recosntr}
J.~Z.~Ma and X.~Zhang,
  %``Probing the dynamics of dark energy with novel parametrizations,''
  Phys.\ Lett.\  B {\bf 699} (2011) 233 [arXiv:1102.2671];
Dragan Huterer, Michael S. Turner  Phys.Rev.D64:123527 (2001)   [astro-ph/0012510];
Jochen Weller (1), Andreas Albrecht, Phys.Rev.D65:103512,2002   [astro-ph/0106079]

\bib{DELind}
M. Chevallier and D. Polarski, Int. J. Mod. Phys. D10,
213 (2001); E. V. Linder, Phys. Rev. Lett. 90, 091301 (2003).

\bib{corasaniti}
P. S. Corasaniti, B. A. Bassett, C. Ungarelli, and E. J. Copeland, Phys. Rev. Lett. 90, 091303 (2003);

\bib{huang}
Z. Huang, J. R. Bond, L.Kofman Astrophys.J.726:64,2011

\bib{tracker}
Steinhardt,P.J.Wang,L.Zlatev I. Phys.Rev.Lett. 82 (1999) 896, arXiv:astro-ph/9807002;
Phys.Rev.D 59(1999) 123504, arXiv:astro-ph/9812313
9.

\bib{quint.ax}
  A.~de la Macorra and G.~Piccinelli,
  %``General scalar fields as quintessence,''
  Phys.\ Rev.\  D {\bf 61}, 123503 (2000)
  [arXiv:hep-ph/9909459];
  %%CITATION = PHRVA,D61,123503;%%
  A.~de la Macorra and C.~Stephan-Otto,
  %``Quintessence restrictions on negative power and condensate potentials,''
  Phys.\ Rev.\  D {\bf 65}, 083520 (2002)
  [arXiv:astro-ph/0110460].
  %%CITATION = PHRVA,D65,083520;%%

\bib{GDE.ax}
  A.~de la Macorra,
  %``A Realistic particle physics dark energy model,''
  Phys.\ Rev.\  D {\bf 72}, 043508 (2005)
  [arXiv:astro-ph/0409523];
  %%CITATION = PHRVA,D72,043508;%%
  A.~De la Macorra,
  %``Quintessence unification models from nonAbelian gauge dynamics,''
  JHEP {\bf 0301}, 033 (2003)
  [arXiv:hep-ph/0111292];
  %%CITATION = JHEPA,0301,033;%%
   A.~de la Macorra and C.~Stephan-Otto,
  %``Natural quintessence with gauge coupling unification,''
  Phys.\ Rev.\ Lett.\  {\bf 87}, 271301 (2001)
  [arXiv:astro-ph/0106316];

\bib{GDM.ax}
  A.~de la Macorra,
  %``Dark group: dark energy and dark matter,''
  Phys.\ Lett.\  B {\bf 585}, 17 (2004)
  [arXiv:astro-ph/0212275],A.~de la Macorra,
  %``BDM Dark Matter: CDM with a core profile and a free streaming scale,''
  Astropart.\ Phys.\  {\bf 33}, 195 (2010)
  [arXiv:0908.0571 [astro-ph.CO]].
  %%CITATION = PHLTA,B585,17;%%

\bib{IDE}
 S.~Das, P.~S.~Corasaniti and J.~Khoury,Phys.  Rev. D { 73},
083509 (2006), arXiv:astro-ph/0510628;
A.~de la Macorra,  %``The Fate of the Universe: Dark Energy Dilution?,''
 Phys.Rev.D76, 027301 (2007), arXiv:astro-ph/0701635

\bib{IDE.ax}
  A.~de la Macorra,
  %``Interacting dark energy: Generic cosmological evolution for two scalar
  %fields,''
  JCAP {\bf 0801}, 030 (2008)
  [arXiv:astro-ph/0703702];
  %%CITATION = JCAPA,0801,030;%%
  A.~de la Macorra,
  %``Interacting Dark Energy: Decay into Fermions,''
  Astropart.\ Phys.\  {\bf 28}, 196 (2007)
  [arXiv:astro-ph/0702239].
  %%CITATION = APHYE,28,196;%%

\bib{ma} C. Ma and E. Bertschinger, Astrophys. J. 455, 7 (1995);

\bib{mukhanov}
J. Garriga and V. Mukhanov, Phys. Lett. B 458, 219 (1999)

\bib{hu}
W. Hu, D. Scott, N. Sugiyama and M. J. White, Phys. Rev. D 52,
5498 (1995), e-Print: astro-ph/9505043. W. Hu, Astrophys. J. 506
(1998) 485 [arXiv:astro-ph/9801234]. S. Bashinsky and U. Seljak,
Phys. Rev. D 69, 083002 (2004); S. Bashinsky, astro-ph/0411013.

\bib{bean}
R. Bean and O. Dore, Phys. Rev. D 69, 083503 (2004).

}

\end{document}